\documentclass[pre,twocolumn,showpacs,preprintnumbers,amsmath,amssymb,superscriptaddress]{revtex4}

\usepackage{graphicx}
\usepackage{dcolumn}
\usepackage{bm}

\newcommand{\kb}{k_\mathrm{B}}

\newcommand{\C}{\tilde C(\omega)}
\newcommand{\R}{\tilde R(\omega)}

\newcommand{\reR}{\tilde R^\prime(\omega)}
\newcommand{\I}{\tilde I(\omega)}

\newcommand{\utphysics}{Department of Physics, Graduate School of Science, University of Tokyo, Hongo, Tokyo 113-0033, Japan}
\newcommand{\utkomaba}{Department of Pure and Applied Sciences, University of Tokyo, Komaba, Tokyo 153-8902, Japan}

\begin{document}
 
\preprint{}

\title{Experimental Test of a New Equality:\\ Measuring Heat Dissipation in an Optically Driven Colloidal System}

\author{Shoichi Toyabe}
\email{shoichi@daisy.phys.s.u-tokyo.ac.jp}
\affiliation{\utphysics}
\author{Hong-Ren Jiang}
\affiliation{\utphysics}
\author{Takenobu Nakamura}
\affiliation{\utkomaba}
\author{Yoshihiro Murayama}
\affiliation{\utphysics}
\author{Masaki Sano}
\email{sano@phys.s.u-tokyo.ac.jp}
\affiliation{\utphysics}

\date{\today}

\begin{abstract}
Measurement of energy dissipation in small nonequilibrium systems is generally a difficult task.
Recently, Harada and Sasa [Phys.\,Rev.\,Lett.\,{\bf 95}, 130602(2005)] derived an equality relating 
the energy dissipation rate to experimentally accessible quantities in nonequilibrium steady states described by the Langevin equation.
Here, we show the first experimental test of this new relation in an optically driven colloidal system.
We find that this equality is validated to a fairly good extent, thus the irreversible work of a small system is estimated from readily obtainable quantities.
\end{abstract}

                           
                             
\maketitle

Experimental techniques of the manipulation and observation in small systems has developed significantly in these years. 
Recently, new relations in statistical mechanics such as the fluctuation theorems\cite{Evans1993,Evans1994,Gallavotti1995,Crooks1999} and the Jarzynski equality\cite{Jarzynski1997,Hatano2001} have been discovered for small systems driven out of equilibrium.
These new ideas have been experimentally tested\cite{Wang2002,Wang2005,Douarche2005,Trepagnier2004,Carberry2004} and utilized\cite{Bustamante2005,Collin2005,Liphardt2002} mainly in the systems of macromolecules and colloids using the optical tweezer.


Recently, Harada and Sasa derived an equality\cite{Harada2005,Harada2006,Teramoto2005} relating the rate 
of energy dissipation to experimentally accessible quantities described by Langevin equations.
Although the energy dissipation rate is the fundamental quantity to characterize the nonequilibrium 
steady states, it is generally hard to measure it directly in small systems, especially in microrheological systems and biological systems such as molecular machines.
The important aspect of this relation is that it enables us to evaluate the energy dissipation rate
 from experimentally accessible quantities such as correlation functions and response functions of fluctuating observables
.
Moreover, this equality was generalized to systems with a memory kernel\cite{Deutsch2006}.
In those systems, the fluctuating force has a coloured spectrum, not a white one.
Such a generalization provides us applications of the equality to much wider systems, including rheological systems and biological systems.
In this paper, we showed the first experimental test of the prediction by Harada and Sasa in a colloidal system driven by an optical tweezer in a viscous fluid.
Also, we evaluated the energy dissipation rate as a function of frequency using this equality.

Firstly, we introduce this equality.
In the over damped case, the Langevin equation is written as
\begin{equation*}
\label{eq:langevin}
\gamma\dot x(t)=F(x(t),t)+ \varepsilon f^p(t) + \hat\xi(t),
\end{equation*}
where $x(t)$ denotes a certain degree of freedom such as the position of a Brownian particle.
The force $F(x(t),t)$ acting on the particle at $x(t)$, can be periodically or randomly changing in time, or time independent in general.
$\varepsilon f^p(t)$ is a small probe force to measure the response and $\hat\xi(t)$ is a white Gaussian random force : $\langle\hat\xi(t)\rangle=0$, $\langle\hat\xi(t)\hat\xi(s)\rangle=2\gamma \kb T\delta(t-s)$.
In the systems described by this Langevin equation, Harada and Sasa derived the following equality\cite{Harada2005,Harada2006,Teramoto2005} , which relates the ensemble average of the energy dissipation rate $J(t)$ to experimentally accessible quantities as :
\begin{equation}
\label{eq:Harada-Sasa}
\left<J\right>_0 = \gamma\int^\infty_{-\infty}\left[\C-2\kb T\reR\right]\frac{d\omega}{2\pi}, 
\end{equation}
where Fourier transform of an arbitrary function $A(t)$ is defined as $\tilde A(\omega)\equiv\int^\infty_{-\infty}A(t)\exp(i\omega t)dt$.
$\tilde A^\prime(\omega)$ denotes the real part of $\tilde A(\omega)$. 
$C(t)$ is the auto correlation function of the velocity :
${\displaystyle C(t)\equiv\left<\dot x(t)\dot x(0)\right>_0.}$
$R(t)$ is the linear response function of the velocity to a small external probe force $\varepsilon f^p(t)$ :
\begin{equation}
\left<\dot x(t)\right>_\varepsilon-v_s = \varepsilon\int^t_{-\infty}\!\!\!\!R(t-s)f^p(s)ds+o(\varepsilon^2),
\end{equation}
where $\left<\cdots\right>_\varepsilon$ denotes an ensemble average under the probe force $f^p(t)$ of order $\varepsilon$.
$v_s$ is the steady state velocity.
Measurement of $\C$ and $\reR$ is usually more accessible by experiments than that of $J(t)$.
The right hand side of Eq.\,(\ref{eq:Harada-Sasa}) vanishes near the equilibrium state due to 
the fluctuation dissipation relation of the first kind\cite{Kubo1991}; $\C = 2\kb T\reR$.
While it has a finite value in nonequilibrium states generally.
Therefore the integral in Eq.\,(\ref{eq:Harada-Sasa}) corresponds to the extent of the FDR violation.

We tested this equality(Eq.\,(\ref{eq:Harada-Sasa})) in an optically driven system, in which one small particle was trapped by an optical tweezer.
The optical tweezer enables us to create a potential for small particles\cite{Neuman2004}.
Nonequilibrium steady states(NESS) can be established by introducing persistent forcing to the
Brownian particle. 

\begin{figure}[htbp]
\includegraphics[width=0.42\textwidth]{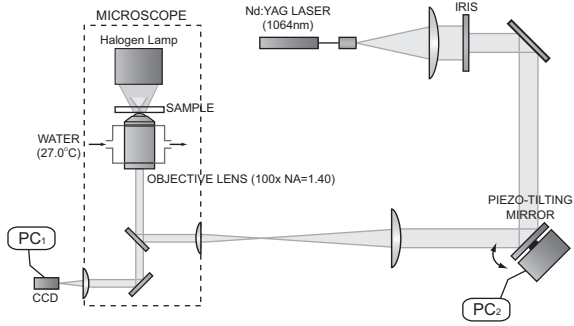}
\caption{Experimental Setup}
\label{fig:Method:Optics}
\end{figure}

The experimental setup is shown in Fig.\,\ref{fig:Method:Optics}.
Infrared laser(Spectra Physics,1064nm) was introduced to the microscope(Olympus, IX71) and focused by an objective lens(Olympus UPLSAPO,100x NA=1.4) to create an optical tweezer.
One carboxylated polystyrene particle of a diameter of 0.984$\pm$0.023$\mu$m(Polysciences,\#08226) was trapped in distilled water.
Trajectory of the particle was captured by a CCD camera(Hamamatsu, HISCA C6770) with 472[frames/s] and analyzed to know the particle position.
The temperature was kept at 27.0$\pm$0.1[$^\circ$C] by water circulation around the objective lens.

\begin{figure}[htbp]
\includegraphics[width=0.35\textwidth]{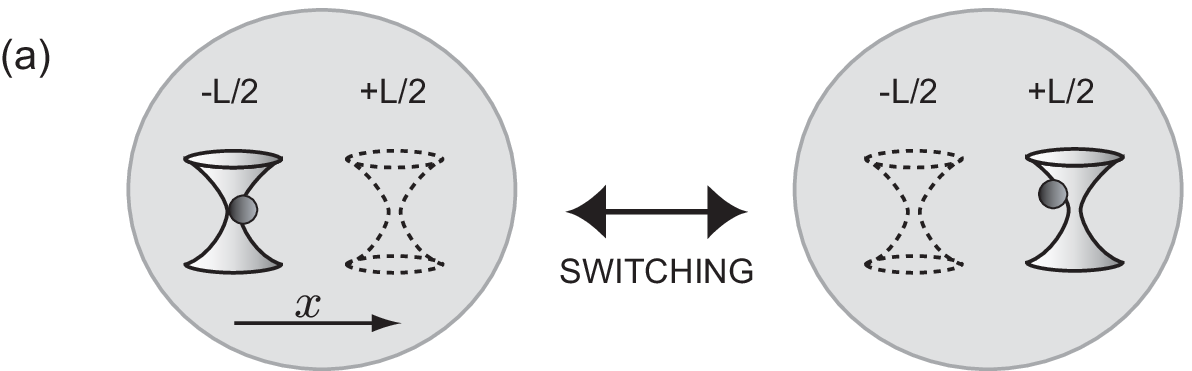}\\
\includegraphics[width=0.39\textwidth]{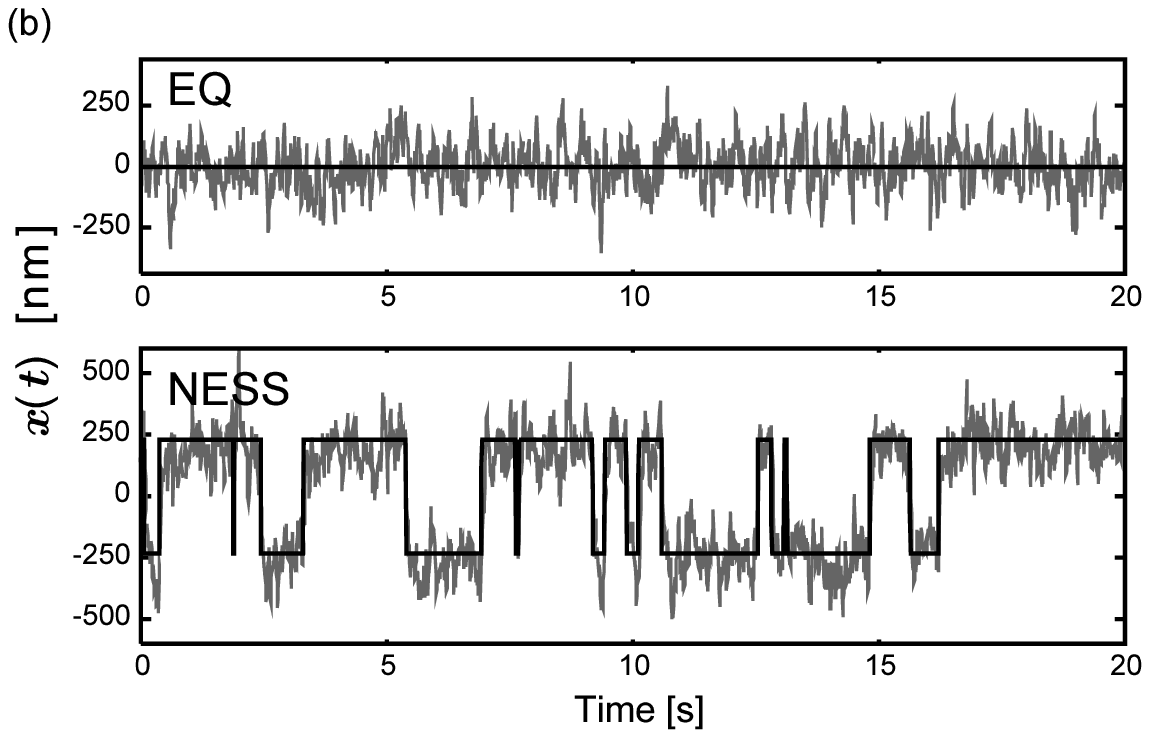}
\caption{Switching: (a)To achieve nonequilibrium steady states(NESS), position of the optical tweezer was switched between two sites($\pm L/2$) using a Piezo mounted mirror at random with Poissonian process. 
In the equilibrium state(EQ), the position of the tweezer was fixed. In NESS, we decided whether to switch or not at random every 4ms. 
One switching takes 5[ms] due to the mechanical response of the mirror unit.
(b)Typical trajectory for EQ and NESS($\lambda=1.0$). 
Black line is the trajectory of the optical tweezer. 
Gray line is the trajectory of the particle.
$L=460.8\pm0.14\mathrm{[nm]}$.}
\label{fig:Method:Switching}
\end{figure}

To achieve nonequilibrium steady states(NESS), we switched the position of the tweezer between two 
sites $\pm L/2$($L=460.8\pm 0.14$[nm]) at random according to the Poissonian process by tilting 
the angle of a Piezo mounted mirror(PI, S-226) controlled by a PC.
The corresponding Langevin equation becomes
\begin{equation*}
\gamma\dot x(t)=-k \left[x(t)-\frac L2\hat\sigma(t)\right]+\varepsilon f^p(t)+\hat\xi(t),
\end{equation*}
where $k$ is the stiffness of the tweezer. 
$x(t)$ is the horizontal coordinate of the particle position (see Fig.\,\ref{fig:Method:Switching}(a)).
$\hat\sigma(t)=\{-1,1\}$ is a random variable obeying Poissonian process with a transition rate $\lambda$;\,$\left<\hat\sigma(t)\hat\sigma(s)\right>=\exp(-2\lambda|t-s|)$.
The particle is sufficiently small($m/\gamma\sim10^{-7}[s]$), thus the over damped assumption is supposed to be valid.

At first, we trapped a particle with fixing the position of the tweezer and obtained a histogram of the particle location(1,703,884 points).
Then, according to the Boltzmann distribution, we calculated the potential energy of the particle, which was well fitted by a harmonic potential with a spring constant of $k=117.5\pm 0.48\mathrm{[k_BT/\mu m^2]}=0.4870\pm 0.0020\mathrm{[pN/\mu m]}$.

\begin{figure}[htbp]
\includegraphics[width=0.39\textwidth]{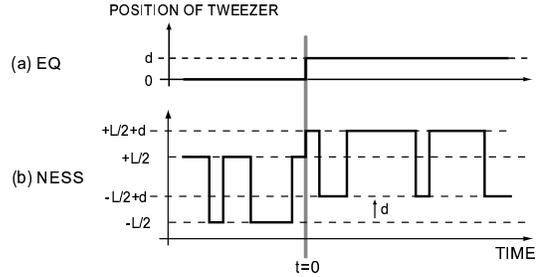}
\caption{Measurement of Response Function\quad :\quad (a)We shifted the tweezer position with a distance of $d=190.2\pm$0.11[nm] at $t=0$ and observed the relaxation processes to the new laser position. 
We repeated this and obtained an ensemble average.
(b)Similarly in NESSs, we added a bias $d$ and observed the relaxation processes.}
\label{fig:Method:R}
\end{figure}

\begin{figure}[htbp] 
\includegraphics[width=0.41\textwidth]{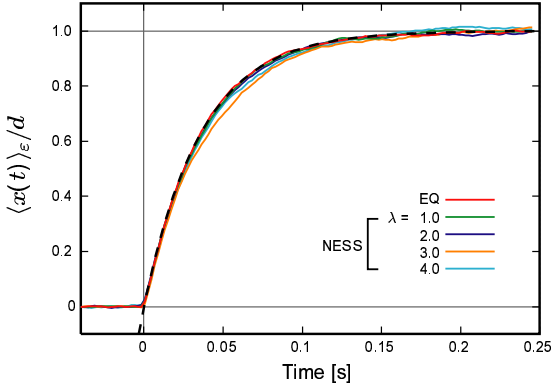}
\caption{(Color) Response Function : The trajectories of the relaxation processes in EQ(equilibrium state(no switching)) and NESSs(nonequilibrium steady states) with mean switching rates $\lambda$= 1.0, 2.0, 3.0, and 4.0[Hz].
Dashed line is the fitting curve : $\left<x(t)\right>/d=1-\exp(-\omega_c t)$ for EQ. $d$ was 
190.2$\pm$0.11[nm] and the cutoff frequency $\omega_c$ was 27.61$\pm$0.05[rad/sec]. 
We averaged 7,854 trajectories for EQ, $\lambda=1.0, 3.0, $ and 4.0 respectively and 8,085 trajectories for $\lambda=2.0$.}
\label{fig:Result:R}
\end{figure}

To know the response function $\R$, we shifted the position of the tweezer at $t=0$ with a distance of $d=$190.2$\pm$0.11[nm]\,(Fig.\,\ref{fig:Method:R}) and tracked particle trajectories in the relaxation processes in the equilibrium state(EQ) and four NESSs(the mean switching rates were varied from $\lambda$=1.0 to 4.0 with a step of 1.0[Hz]).
The corresponding probe force $\varepsilon f^p(t)$ becomes $kd\,(t\ge 0)$, $0\,(t<0)$.
Figure\,\,\ref{fig:Result:R} shows that the ensemble averaged trajectories of the relaxation processes in EQ and NESSs coincided well. 
We fitted an exponential function $\left<x(t)\right>/d=1-\exp(-\omega_c t)$ to the trajectory in EQ. 
The cutoff frequency was estimated as $\omega_c$=27.61$\pm$ 0.05[rad/sec], which yields the friction coefficient
 $\gamma= k/\omega_c=1.763\times 10^{-8}$[kg/s]\cite{FaxenMemo}.
%

From the fitting curve in Fig.\,\ref{fig:Result:R}, we obtained the response function $\reR=\omega^2/(\omega^2+\omega_c^2)\gamma$.
We compared this response function $\reR$ and correlation functions of the velocity $\C$ obtained from the trajectories of the particle.
\begin{figure}[htbp]
\includegraphics[width=0.39\textwidth]{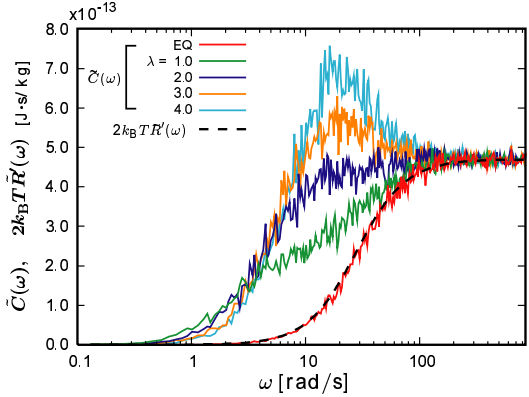}
\caption{(Color) $\C$ and $2\kb T\reR$ : Dashed line is the response function $2\kb T\reR=2\kb T\omega^2/(\omega^2+\omega_c^2)\gamma$.
52[runs]$\times$69.17[s] (1,703,884[points]) in EQ and NESS at $\lambda=3.0$, 4.0 respectively and 53[runs]$\times$69.17[s] (1,736,651[points]) in NESS at  $\lambda=1.0$ and $2.0$ respectively were averaged for $\C$.  
Moving averages were taken in $\omega$ direction.}
\label{fig:Result:C}
\end{figure}
In Fig.\,\,\ref{fig:Result:C}, we found a good coincidence between $\C$ and $2\kb T\reR$ in the equilibrium state.
Thus, the fluctuation dissipation relation was confirmed.
While, these had differences in nonequilibrium steady states.
Also, the higher the switching rate $\lambda$ was, the larger the deviation of $\C$ from $2\kb T\reR$ was.
\begin{figure}[htbp]
\includegraphics[width=0.42\textwidth]{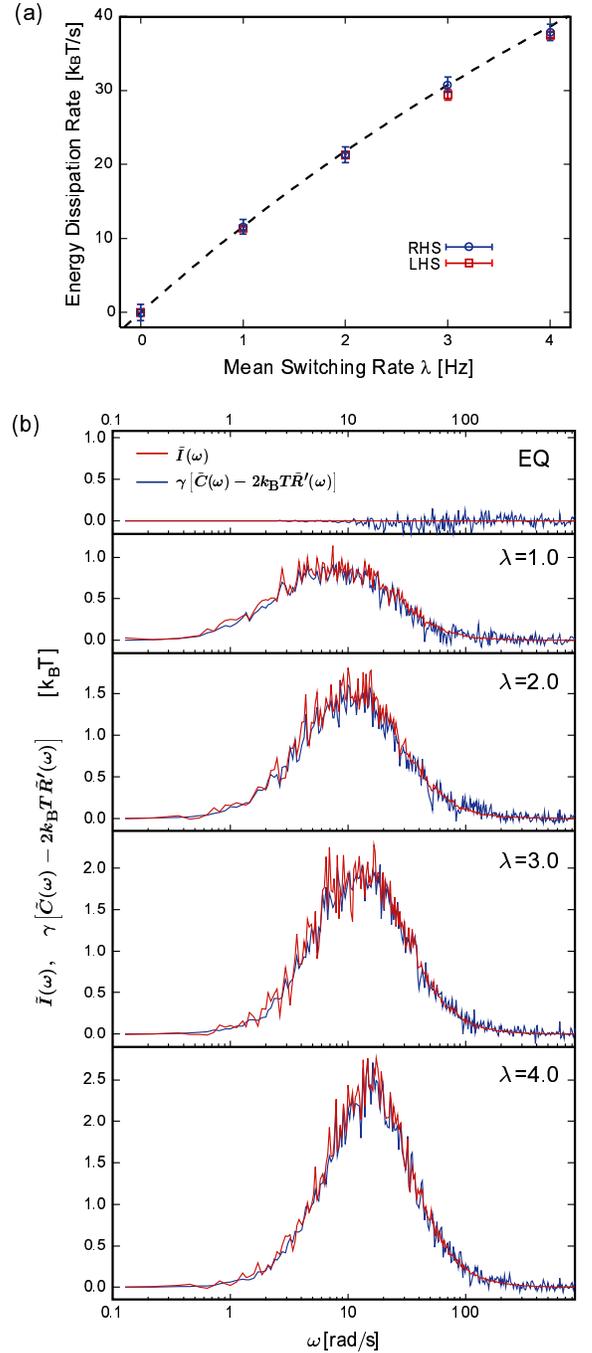}
\caption{(Color) (a)The left hand side(LHS) and the right hand side(RHS) of the equality(\ref{eq:Harada-Sasa}) were plotted against the mean switching rate $\lambda$[Hz]. 
For RHS, we integrated only in the region $\omega<850$  because the difference between LHS and RHS should vanish at sufficiently high frequencies.
Dashed line is the theoretical curve\,\cite{Harada2006-2}.
The standard errors of RHS were around 1[k$_\mathrm{B}$T/s] at all switching rates.
(b)Validation of the equality for each $\omega$(Eq. \ref{eq:I(w)}); $\I$\,(blue line) vs $\gamma[\C-2\kb T\reR]$ 
(black line).  We used the same data for $\C$ and $\reR$ in Fig.\,\ref{fig:Result:C} to calculate $\gamma[\C-2\kb T\reR]$}.
\label{fig:Result:JI}
\end{figure}

Equation\,\,(\ref{eq:Harada-Sasa}) implies that the product of $\gamma$ and the area between $\C$ and $2\kb T\reR$ in Fig.\,\ref{fig:Result:C} corresponds to the energy dissipation rate.
The irreversible energy dissipation should be balanced with the energy input in steady states, thus 
 $\langle J\rangle=\langle F(x(t),t)\circ v(t)\rangle$, where $\circ$ denotes the Stratonovich multiplication\,\cite{Sekimoto1997}. 
On the contrary, in our setup it is possible to evaluate this $\langle J\rangle$ because the force exerted by the optical tweezer can be calculated from the displacement between the particle and the center of the tweezer as $F(x(t),t)=-k(x(t)-\frac L2\sigma(t))$\,\, 
 (Note that this is not always possible in other systems.). 
We compared these two values in Fig.\,\ref{fig:Result:JI}(a) and confirmed that the equality is valid to good extent within statistical error.
In one switching, the particle is supposed to travel a distance of $L$ on an average.
This corresponds to an energy injection of $kL^2/2=12.5$[k$_\mathrm{B}$T] per one switching and $\lambda kL^2/2=12.5\lambda$[k$_\mathrm{B}$T] per unit time.
However, insufficient relaxation to the switched position due to fast switchings causes a deviation from this value.
The dashed line in Fig.\,\ref{fig:Result:JI}(a) is the theoretical value including this effect\cite{Harada2006-2}.

It is possible to derive a local equality(Eq.\,(\ref{eq:Harada-Sasa})):
\begin{equation}
\label{eq:I(w)}
\I=\gamma\left[\C-2\kb T\reR\right],
\end{equation}
where $\I$ is the Fourier transform of the cross correlation of the force $F(x(t),t)$ and velocity
 $v(t)$ at different times:
\begin{equation*}
I(t)\equiv \left[\left<F(x(t),t)\circ v(0)\right>_0+\left<F(x(0),0)\circ v(t)\right>_0\right]/2. 
\end{equation*}
Equation\,\,(\ref{eq:I(w)}) implies that the equality holds at each $\omega$\cite{Harada2006}.
This local relation provides us a chance for the more detailed experimental test of the equality.
Figure\,\,\ref{fig:Result:JI}(b) shows a good agreement of the both sides of Eq.\,(\ref{eq:I(w)}) at each $\omega$.
As we increased the mean switching rate $\lambda$, the frequency and height of the peak of the local dissipation rate increased.

We showed the first experimental demonstration of the Harada and Sasa's equality.
Using this equality, we evaluated the rate of the energy dissipation for an optically driven particle 
with a precision of around 1[k$_\mathrm{B}$T/s] from the correlation function and response function 
of the particle velocity.
Note that this equality is not applicable to systems with retarded friction, since it is based on a Markovnian Langevein equation. 
However, systems which are more interesting and practical often have a retarded friction. 
For example, molecular machines are working in extremely crowded environment in cells. 
In such a situation, non-Markovnian behaviour is expected. 
Also, microrheology is attracting extensive studies recently.
As we described, Deutsch and Narayan recently generalized the equality to these situations, namely systems described by a generalized Langevin equation\cite{Deutsch2006}.
If this generalization is confirmed by experiments, it provides us applications of this equality to much wider systems.
This will be the primary focus of the forthcoming experiments.

\begin{acknowledgments}
We appreciate helpful suggestions of Prof.\,Takao Ohta, Prof.\,Shin-ichi Sasa, and Dr.\,Takahiro Harada.
This work was supported by the Ministry of Education, Science, Sports, and Culture of Japan (No.16206020, No.18068005).
\end{acknowledgments}
\bibliography{article004}

\end{document}